\documentclass[preprint2]{aastex}

\usepackage{epsfig}

\newcommand{\etal}{{\it et al.\ }}

\newcommand{\avg}[1]{\langle{#1}\rangle}
\newcommand{\abs}[1]{\mid{#1}\mid}

\newcommand{\ltsima}{$\; \buildrel < \over \sim \;$}
\newcommand{\lsim}{\lower.5ex\hbox{\ltsima}}
\newcommand{\gtsima}{$\; \buildrel > \over \sim \;$}
\newcommand{\gsim}{\lower.5ex\hbox{\gtsima}}

\def\gtrsim{\mathrel{\hbox{\rlap{\hbox{\lower4pt\hbox{$\sim$}}}\hbox{$>$}}}}

\def\lesssim{\mathrel{\hbox{\rlap{\hbox{\lower4pt\hbox{$\sim$}}}\hbox{$<$}}}}

\begin{document}
\title{Fast Clustering Analysis of Inhomogenous Megapixel CMB Maps}

\author{Istv\'an Szapudi,\footnote{ Institute for Astronomy, 
University of Hawaii, 2680 
Woodlawn Dr, Honolulu, HI 96822} Simon Prunet,\footnote{ CITA, University 
of Toronto, 60 St George St, Toronto, Ontario, M5S 3H8, Canada}$^{,4}$
Stephane Colombi\footnote{ Institut d'Astrophysique de Paris, CNRS, 
98bis bd Arago, F-75014 Paris, France}$^,$\footnote{ NIC 
(Numerical Investigations in Cosmology), CNRS} 
}

\begin{abstract}
Szapudi \etal (2001) introduced the method of estimating
angular power spectrum of the CMB sky via 
heuristically weighted correlation functions.
Part of the new technique is that all (co)variances are evaluated
by massive Monte Carlo simulations, therefore a fast way
to measure correlation functions in a high resolution map is essential.
This letter presents a new algorithm to calculate pixel space
correlation functions via fast spherical harmonics transforms.
Our present implementation of the idea
extracts correlations from a MAP-like CMB map 
(HEALPix resolution of $512$, i.e. $ \simeq 3 \times 10^6$ pixels) in about 
5 minutes on a 500MHz computer, including $C_\ell$ inversion;
the analysis of one Planck-like map takes less then one hour.
We use heuristic window and noise weighting in pixel space, 
and include the possibility of additional signal weighting as well,
either in $\ell$ or pixel space.
We apply the new code to an ensemble of MAP simulations,
to test the response of our method 
to the inhomogenous sky coverage/noise of MAP.
We show that the resulting $C_\ell$'s 
are very close to the theoretical expectations.
The HEALPix based implementation of the method, SpICE (Spatially
Inhomogenous Correlation Estimator) will be available to the
public from the authors.

\end{abstract}


\vfill


\section{Introduction}

With the successful launch of MAP and the advancing 
Planck schedule, megapixel Cosmic Microwave 
Background (CMB) maps are just around the corner.
To fully realize their potential of
constraining cosmological parameters within
a few percent (e.g., Spergel 1994; Knox 1995; Hinshaw, Bennett, \& Kogut
1995; Jungman \etal 1996; Zaldarriaga, Spergel \& Seljak 1997; Bond, 
Efstathiou, \& Tegmark 1997), data analysis methods have
to face hitherto unprecedented challenges. 
The standard maximum likelihood developed for COBE
(e.g., G\'orski 1994; G\'orski \etal 1994, 1996; Bond 1995; Tegmark
\& Bunn 1995; Tegmark 1996; Bunn \& White 1996;
Bond, Jaffe, \& Knox 1998, 2000) is already
pushing supercomputers for balloon-borne and ground based 
experiments with $N \simeq 10^5$ pixels 
(de Bernardis \etal 2000; Netterfield \etal 2001;
Hanany \etal 2000; Jaffe \etal 2000; Martin \etal 1996; 
Miller \etal 1999; Peterson \etal 2000), and it is clearly
inadequate for analysing megapixel maps on any future supercomputer.
This letter addresses an important step in the full data processing
pipeline from component separation/mapmaking to 
cosmological parameter estimation: the fast estimation of the angular power
spectrum, $C_\ell$ from megapixel map with realistic
geometry (sky cut, cut out holes), and inhomogeneous
coverage and noise. 

A recent surge of activity motivated by the obvious need for
fast alternatives to the standard maximum likelihood estimation
produced an array of promising techniques. Several of them
use symmetries of a particular experiment to gain speed.
An experiment specific technique for MAP was developed by
Oh, Spergel and Hinshaw (1998) making use of the
planned approximate azimuthal symmetry
of the coverage/noise. Similar developments are under way for Planck
(e.g., Wandelt 2000) possibly exploiting symmetries in its scanning strategy.
Other techniques give up optimality in favor of speed.
Correlation functions with simple weights were advocated
by Szapudi \etal (2001, hereafter SPPSB), while an analogous method
based on direct spherical Fourier transform was developed
by Hivon \etal (2001). In the standard CMB formalism
(e.g. Bond \etal 1998) both of these can be thought of 
as quadratic estimators with suboptimal weights, yet,
they tend to produce results very close to optimal.

This letter presents significant improvements of the
correlation function method. We present a fast algorithm
for extracting the pixel space estimator using fast spherical harmonics 
transform,  introduce noise
weighting, test the use of Monte Carlo realizations
to calculate noise correlation functions in the
estimator, and finally illustrate the possibility of additional
signal weighting. 

\begin{figure} 
[htb!]
\centerline{\hbox{\epsfig{figure=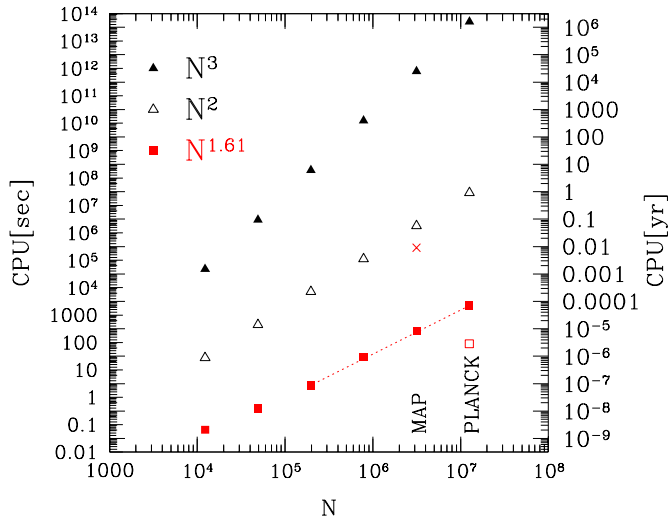,width=9cm}}}
\caption{CPU requirements of selected CMB
analysis methods are plotted: MADCAP ($N^3$, filled triangles,
Borrill 1999),
SPPSB ($N^2$, open triangles), and the present
method ($\simeq N^{1.61}$, filled squares).
The experiment specific technique of OSH for MAP is plotted with a cross,
while an open square illustrates how Moore's law will shift the
CPU for our method by 2007. All the squares are actual
measured values on a 500Mhz CPU,
while the triangles are based on extrapolation of scaling.
The two leftmost points are relevant for MAP
and Planck. The righthand $y$-axis displays CPU time in years, instead
of seconds for clarity.
}
\label{fig3}
\end{figure}       

\section{Summary of the Method}

Our latest recipe to extract $C_\ell$'s from
a large CMB map is an improved version of
SPPSB: extract
the two-point correlation function with an unbiased weighted
estimator sampled at the roots of Legendre polynomials, 
then integrate with a Gauss-Legendre quadrature to obtain the $C_\ell$'s.
Next we review the correlation function
estimator.

Let us denote the temperature fluctuations at a sky vector
$n$, a unit vector pointing to a pixel on the sky, with
$T(n)$. In isotropic universes the two-point correlation
function is a function only of the angle between the
two vectors and can expanded into a Legendre series,
\begin{equation}
  \xi_{12} = \avg{T(n_1)T(n_2)} = 
  \sum_\ell \frac{\ell + 1/2}{\ell ( \ell+1)} {\cal C}_\ell P_\ell(\cos\theta),
  \label{eq:xi}
\end{equation}
where $\cos\theta = n_1 . n_2$ is the dot product of the two unit vectors,
$P_\ell(x)$ is the $\ell$-th Legendre polynomial, and the 
${\cal C}_\ell$ coefficients
realize the angular power spectrum of fluctuations. If the 
CMB anisotropy is Gaussian, which is expected to be
an excellent approximation, the correlation function,
or the ${\cal C}_\ell$'s, yields full statistical description.

In reality each pixel value, $\Delta_i$, contains contributions
from the CMB and noise; the latter is also assumed to be
Gaussian with a correlation matrix (the noise matrix) $N_{ij}$,
determined during map-making. The full pixel-pixel correlation
matrix is
  $C_{ij} = \xi_{ij}+N_{ij}.$
matrix, as a sum of noise correlation functions measured in
a set of noise realizations. Therefore we use the
modified version of the estimator by SPPSB
\begin{equation}
  \tilde\xi(\cos\theta) = \sum_{ij} f_{ij} (\Delta_i \Delta_j - 
    \frac{1}{M}\sum_{k=1}^M n_i^k n_j^k),
  \label{eq:modestimator}
\end{equation}
where $n_i^k$ is one of $M$ realizations of the noise for pixel $i$.
Our estimator can be extracted efficiently and
accurately for a general noise matrix, 
as long as a fast noise generator exists.
The weights $f_{ij}$ merit further discussion. For $f_{ij} \simeq (S+N)^{-2}$
the above corresponds to an optimal quadratic estimator (where
$S$ denotes the matrix corresponding to $\xi_{ij}$). Unfortunately,
the cost to calculate the optimal weights would be prohibitive
as it is for any other version of the
optimal quadratic estimator; speed is gained by using
heuristic weights instead of the optimal ones. 
SPPSB used  window weighing, $f_{ij} = 0$
unless the pair of pixels belong to a particular
bin in $\cos\theta$, and $\sum_{ij}f_{ij} = 1$.
The analysis of the next section uses more general
noise weighting. Other possibilities exist and are worth
to explore, however, they can only affect slightly the error bars of our
unbiased estimator. 
The next section demonstrates that 
heuristic noise weighting produces results close to the theoretical
expectations.
Heuristic window weighting is natural in
pixel space where the window is diagonal, and it amounts
to edge effect correction (Szapudi \& Szalay 1998). Heuristic
noise weighting takes this idea one step further, and it is
again natural in pixel space as long as the noise is localized
as it is for MAP. It is worth reemphasizing
that azimuthal symmetry is not required.

\begin{figure} 
[htb]
\centerline{
\psfig{figure=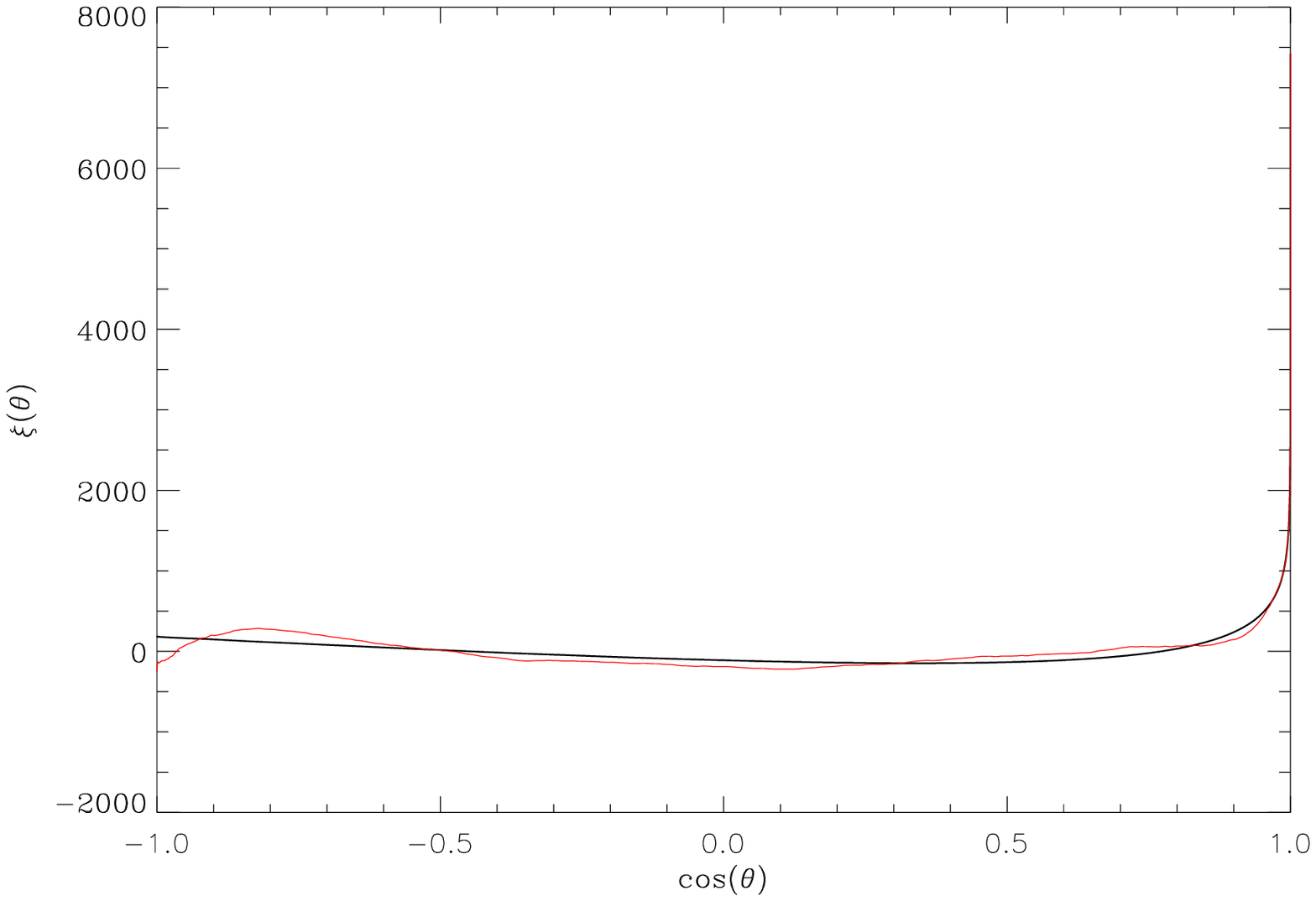,width=8cm}
\hspace{-6.5cm}
\raisebox{1.8cm}{\psfig{figure=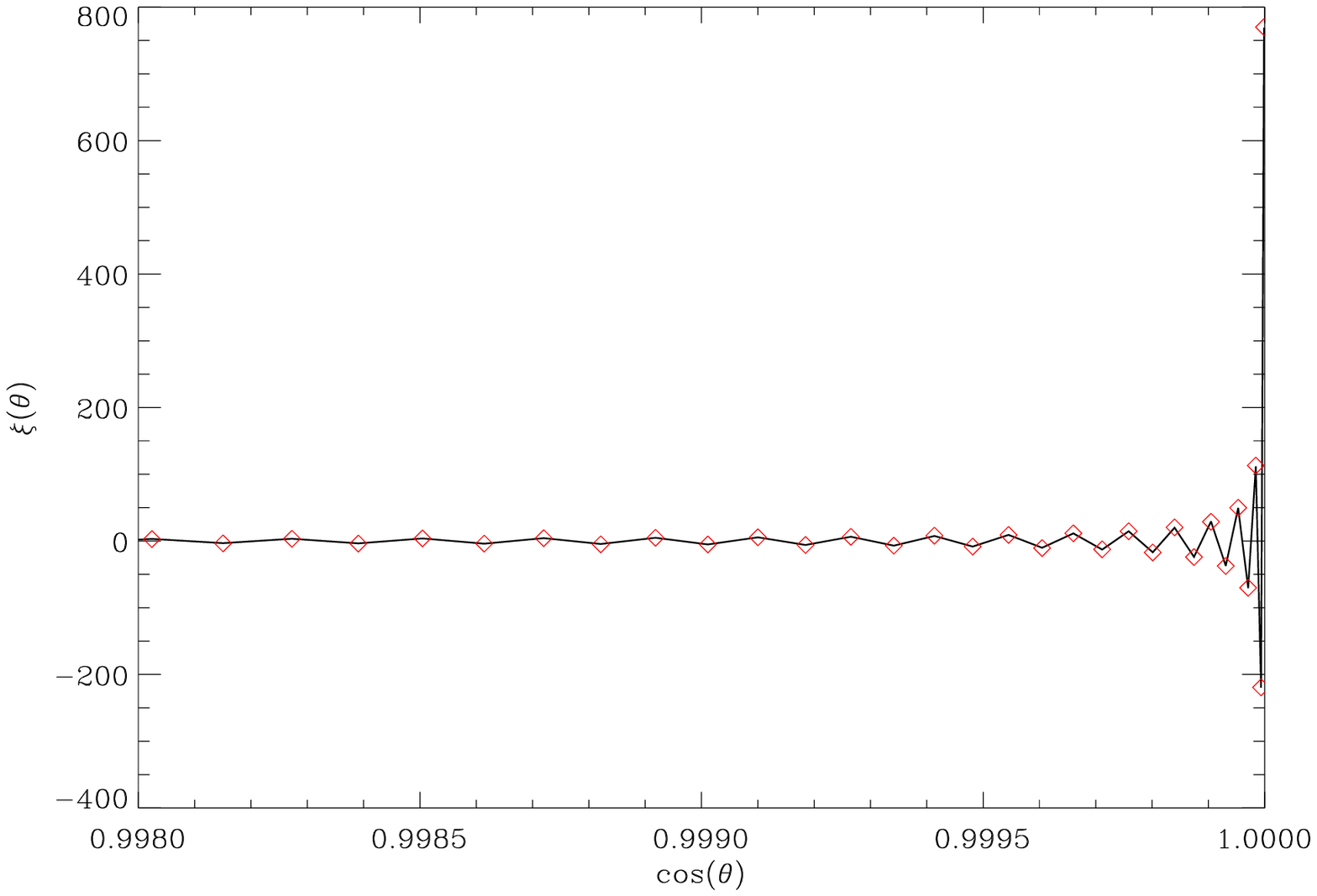,width=5.0cm} }
\hspace{1cm}
}
\caption{
The figure displays the average correlation
function in a simulated MAP survey (black) and one
realization (red). The inserted panel shows the average noise
correlation function (line) with one realization (diamonds).
}
\label{fig5}
\end{figure}       

\section{Fast Estimation of the Correlation Function at
Legendre Roots}

The novel correlation function estimator is based on
fast spherical harmonics decomposition; our aim is to make use of
their speed, based on the FFT's of isolatitude pixel annuli. 
Practical implementation of our estimator is based on the fast
transforms available in the HEALPix
\footnote{\rm http://www.eso.org/kgorski/healpix} package.
Let us define two 
unit vectors with angle $\theta$ apart, e.g.,
$n_1 = (0,0)$, and $n_2 = (\theta,0)$
in spherical coordinates. In the continuum limit,
the raw weighted  pairwise estimator
is an average over the rotation group
\begin{equation}
  D_{f\delta}(cos\theta) = \frac{1}{8\pi^2}\int dg 
   f\delta(R(g)n_1)f\delta(R(g)n_2),
\end{equation}
where the integration is over SO(3), $8\pi^2$ is the full volume
of the Lie group, and R(g) is the transformation corresponding
to the Lie group element. The sum in equation~(\ref{eq:modestimator})
is replaced by an integral, multiplicative weighting is 
assumed: $f_{ij} = f_i f_j$, and $f\delta \simeq f_i \Delta_i$
(both taken at the same vector),
corresponds to the weighted temperature fluctuations  in
the continuum limit.  After a simple calculation based on the
general orthogonality of group representations
\begin{equation}
  D_{f\delta}(cos\theta) = \sum_{lm} \abs{a_{lm}}^2 \frac{1}{4\pi} P_\ell(cos\theta),
\label{eq:fft}
\end{equation}
where the $a_{lm}$'s  are the coefficients of the spherical harmonics
decomposition of $f\delta$.
The ensemble average of the equation yields 
$\avg{D_{f\delta}} = (\xi\ +N) D_{f}$, where we introduced 
the weight correlation function $D_f$, calculated
similarly to $D_{f\delta}$. The
unbiased correlation function estimator can be obtained as
\begin{equation}
  \tilde\xi(cos\theta)
   = \frac{D_{f\delta}(cos\theta)-\bar D_{fn}(cos\theta)}{D_f(cos\theta)},
\end{equation}
where $\bar D_{fn}$ is the average raw noise correlation function calculated
from $M$ simulations;
obviously $\avg{\tilde\xi} = \xi$. According to equation~(\ref{eq:fft})
this estimator is calculated via two spherical harmonics transforms in discrete
pixel space,  with the summation performed up to
$\ell_{max}$. This is a fast realization of  
the estimator in equation~(\ref{eq:modestimator}) with 
multiplicative weights.

We cross-compared the above algorithm with a naive
$N^2$ correlation code, and the results were identical.
To obtain the angular power spectrum, we then
sample the above equation exactly at the roots of
Legendre polynomials, and perform Gauss-Legendre
integration, as done by SPPSB.

The expected scaling is $N^{3/2}{\rm log}N$, corresponding to
an effective $N^{1.61}$ slope for
the present implementation in the range of $N$ we tested.
The speed between calculating the correlation
estimator and the $C_l$ inversion is divided approximately
evenly; on a 500Mhz Dec alpha our present HEALpix compatible
implementation takes $\sim$260 seconds, Planck analysis
is feasible in about 40 minutes. 

The above procedure appears unnatural at first: if we are using
FFT's, and our final goal is to estimate the angular power spectrum,
why do we leave Fourier space at all? The reason is simple:
window and noise are localized in pixel space, thus we can
construct an unbiased estimator with a simple normalization.
Such a normalization in pixel space is equivalent to an edge
effect correction, and heuristic weights can be constructed
intuitively and naturally.
In principle an analogous procedure can be designed
in $\ell$ space, but the non-diagonality of the window
function results in a complex coupling matrix infused with
Wigner $3j$ symbols, as
Hivon \etal (2001) have shown in a tour de force calculation.
The computation and inversion of this matrix to unbias
the estimator is highly non-trivial numerically, especially
in the presence of noise and complex geometry; so far
they have demonstrated numerical feasibility for a relatively simple
ellipsoidal window without any noise weighting. 
It is not inconceivable that this procedure can be 
successfully extended
to more general windows and noise weighting, but at a price
which can be regarded as unnecessary complexity; 
therefore we recommend the simple technique
of constructing an unbiased weighted pixel space estimator instead.

\section{Application: MAP Simulations}

We have generated 1200 MAP simulations 
using HEALPix with inhomogenous sinusoidal coverage assuming a
detector sensitivity of $20\mu K$ for a $0.3^\circ \times 0.3^\circ$
pixel, taken from the MAP homepage, ({\tt http://map.gsfc.nasa.gov/}).
We assumed a Gaussian beam of $18$ arcminutes, we neglected any
sidelobes or asymmetries. Our simulations are similar, although
not perfectly equivalent to that of OSH. In addition to the
noisy MAP simulations, we have generated 1200 pure noise simulations
to be used in our estimator of Equation~(2). 
Our noise weighting was
inversely proportional to the expected noise in a pixel, 
$f_{ij} \simeq 1/(\sigma^p_{n_i} \sigma^p_{n_j})$, with
$p=1$ motivated by prewhitening, and $p=2$ by approximate
minimum variance estimator. Both performed quite similarly,
we show the results from $p=2$.
We used a galactic cut of $\pm 20$ degrees.

The noise correlations were calculated with the code
and were subtracted from the correlation function according
to  equation~(\ref{eq:modestimator}). 
While the noise was assumed
to be diagonal, our practical implementation does
not make use of this property, except in the heuristic
noise weighting we adopted. 
If this assumption
would break, our method would not be affected, as long as
the simulations can be generated quickly. 

\begin{figure} 
[htb]
\centerline{\hbox{\psfig{figure=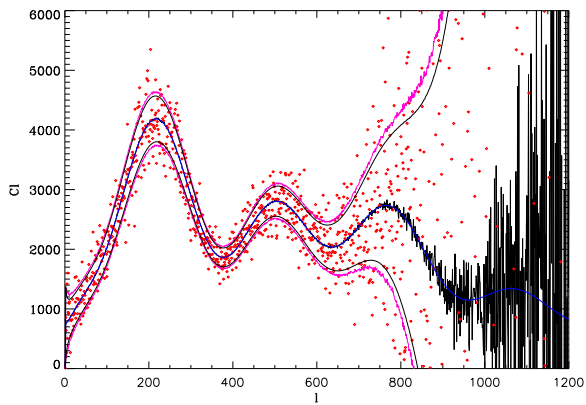,width=8cm} }}
\caption{$C_\ell$ measurements in MAP simulations are illustrated.
The blue line is the underlying
theory, while the overlapping black solid line corresponds the the
average of 1200 measurements; this explicitly demonstrates
the unbiased nature of our technique.
The pair of black lines opening
up for high $\ell$ displays approximate theoretical error bars,
according to equation~(\ref{eq:errorbar2}). The magenta lines
are error bars calculated from 1200 simulations. They track
the theoretical error bars quite well.}
\label{fig:figmap}
\end{figure}       

The $C_\ell$ integration was performed via a Gauss-Legendre quadrature.
Beam and pixel window functions were accounted for as usual.
The results are displayed in Figure \ref{fig:figmap}. Along with
the average $C_\ell$, the theoretical input, as well as theoretical
and measured errors are displayed, the former from the
following approximation (Hivon \etal 2001)
\begin{equation}
\Delta{\cal C}_\ell = \sqrt{\frac{2}{\nu_l}}\left(
{\cal C}_\ell + B(\ell)^2 W(\ell)^2 N(l) \right),
\label{eq:errorbar2}
\end{equation}
where $\nu_l = (2\ell+1) f_{sky} w_2^2/w_4$ with $w_m = \frac{1}{4 \pi}
\int d\Omega f^m(\Omega)$, the $m$-th moments of the weight function $f_i$,
$B(\ell), W(\ell), N(\ell)$ represent beam, pixel, and
noise effects, respectively, 
and $f_{sky}$ is the sky fraction in the map.  According to
Figure~\ref{fig:figmap} 
our method is unbiased, and accurate for MAP: 
the error bars track closely the theoretical expectation within
the accuracy of the approximate theory. The suboptimality
of our technique is expected to be small for MAP-like 
surveys, but this needs to be confirmed by detailed comparison
to optimal techniques. Our figure is similar to Figure 4 of 
Oh \etal (1999), but there are small differences: it appears
that we have assumed a slightly higher noise and/or wider beam. 
It would be worthwhile to perform direct comparison of the two methods
to assess the degree of suboptimality of ours, and the sensitivity
of the experiment specific technique 
to a potential breakdown of the underlying assumptions.

\section{Summary and Discussions}

We have demonstrated that i) correlation functions
with generalized edge correction
can be extracted with full accuracy for a megapixel
map in minutes on a modest personal workstation
ii)
the resulting $C_\ell$ computed from Monte-Carlo realizations
of MAP observations with inhomogeneous coverage agree quite
well with the (approximate) theoretical error bars inferred from 
the moments of the window function applied to the data to construct 
our estimator, as well as the cosmic variance. 
The results should be very close to optimal at high $\ell$
where the noise variance become dominant provided that the noise
matrix is diagonally dominant, but this needs to be confirmed
by a detailed comparison with optimal methods.
The theoretical scaling of our method is $N^{3/2}{\rm log}N$, with a measured
scaling of $\sim N^{1.61}$ in our present implementation,
for the range of $N$ we tested.

In the previous tests we have used MAP-like
diagonal noise/coverage, no other (e.g., azimuthal) symmetries 
were assumed about noise,
or geometry of the map. Cut out holes around bright sources,
or any irregularity in the sampling, represent
only minor perturbations to our method, therefore they should
have no discernible effect on its speed or performance; the
investigation of this point in sufficient detail is left for future
work. Our approach is straightforward to generalize for
non-diagonal noise matrix, specially in the case where the off-diagonal
elements of the noise matrix correspond primarily to pairs of pixels of fixed 
angular separation, as is expected for MAP due to the differential nature
of the measurement.
In the general case, our main assumption is the
fast generation of noise realizations in pixel space, 
which will be supplied by fast mapmaking methods (e.g. Wright 1996; 
Prunet \etal 2001).

At the heart of our method is the use of heuristic 
weighting corresponding to general edge effect correction.
(Indeed, noise can be considered as a fuzzy window, or
window can be considered as infinite noise). The simplicity
of our weighting amounts to a substantial gain in speed over
the costly optimal weighting $(S+N)^{-2}$. In addition,
the use of spherical harmonics decomposition
facilitates heuristic signal weighting, which
is natural in $\ell$ domain.
The sum of the $a_{lm}$'s in equation~(\ref{eq:fft})
can be recognized as a pseudo $C_l$
of the weighted fluctuations estimated with uniform weights in
$\ell$ space.
For low $\ell$'s, for complicated noise and window patterns,
it might be desirable to sum
the $a_{lm}$'s with weights obtained from
the correlation matrix $<a_{l_1m_1}a_{l_2m_2}>$ evaluated from
Monte Carlo simulations. This matrix can 
be diagonalized e.g. up to a certain $\ell$, or within
a band of $\Delta\ell$, to calculate approximate (semi-heuristic)
weights for our estimator of the correlation function.   While such
improvements do not seem necessary for MAP, a clear
upgrade path for our method exists for more complex experiments
if needed. Although more natural in $\ell$ space,
heuristic signal weighting 
is possible in pixel space as well
(e.g., Colombi, Szapudi, \& Szalay 1998).

Our technique has further potential besides speeding up $C_l$ estimation:
it opens up a full array of possible applications
and generalizations. These include cross-correlations
(between channels for component separation, between LSS and CMB,
 B-type polarization and lensing, etc) non-Gaussianity
(e.g. 3-point function/bispectrum, cumulant correlators for
SZ, lensing, etc), vector and tensor correlations (for
polarization).  The idea
of real space window and noise weighting is applicable to power spectrum
estimators of galaxies, clusters, lensing etc. as well. 
A Euclidian version of our algorithm is entirely analogous to the
spherical case. It will be
useful  for fast edge and noise corrected  estimation of the power spectrum, 
bispectrum, $N$-point
correlation function and cumulant correlators in galaxy
catalogs.  These generalizations presently under implementation
will be discussed in subsequent papers and
included in a later version of SpICE. 

We would like to thank Dick Bond, Dmitry Pogosyan, Alex
Szalay for their help, and 
Carlo Contaldi and Eric Hivon for useful discussions.
This research was supported by NASA through 
the Applied Information Systems Research program (NAG5-10750).


\begin{thebibliography}{}

\bibitem[]{} Bond J.R., 1995, \prl, 74, 4369 
\bibitem[]{} Bond, J.R., Efstathiou, G., \& Tegmark, M. 1997, \mnras, 291, L33 
\bibitem[]{} Bond, J.R., Jaffe, A.H. \& Knox, L. 1998, \prd, 57, 2117 
\bibitem[]{} Bond, J.R., Jaffe, A.H. \& Knox, L. 2000, \apj, in press
\bibitem[]{} Borrill, J., 1999, Proc. of the 5th European SGI/Cray MPP
\bibitem[]{} Bunn, E.F, \& White, M. 1997, \apj, 480, 6 
\bibitem[]{} Colombi S., Szapudi I., Szalay A.S., 1998, MNRAS, 296, 253
\bibitem[]{} de Bernardis, \etal 2000, Nature, 404, 955 
\bibitem[]{} G\'orski, K.M. 1994, \apj, 430, L85
\bibitem[]{} G\'orski, K.M \etal \& 1994, \apj, 430, L89 
\bibitem[]{} G\'orski, K.M \etal \& 1996, \apj, 464, L11 
\bibitem[]{} Hanany, S. \etal, 2000, \apjl, submitted (astro-ph/0005123)
\bibitem[]{} Hinshaw, G., Bennett, C.L., \& Kogut, A. 1995, \apj, 441, L1 
\bibitem[]{} Hivon, E., \etal 2001, \apj, submitted (astro-ph/01055302)
\bibitem[]{} Jaffe, A.H., etal, 2001, \prl  86, 3475
\bibitem[]{} Jungman, G., Kaminonkowski, M., Kosowski, A., \& Spergel, D.N. 
1996, \prd, 54, 1332 
\bibitem[]{} Knox, L. 1995, \prd, 52, 4307 
\bibitem[]{} Martin, N., \etal 1996, in Space Telescopes and Instruments IV 
Proc.  SPIE (eds. P. Y. Bely and J. B. Breckinridge), 2807, 86 
\bibitem[]{} Miller, A.D., \etal 1999, \apj, 524, 1 
\bibitem[]{} Netterfield, C.B. \etal 2001, \apj, submitted (astro-ph/0104460) 
\bibitem[]{} Oh, S.P., Spergel, D.N., \& Hinshaw, G. 1999, \apj, 510, 551 
\bibitem[]{} Peterson, J.B., \etal 2000, \apj, 532, 83 
\bibitem[]{} Prunet, S., \etal 2001, astro-ph/0101073
\bibitem[]{} Tegmark, M. 1996, \apj, 464, L35 
\bibitem[]{} Tegmark, M. \& Bunn, E.F. 1995, \apj, 455, 1 
\bibitem[]{} Smoot \etal 1992, \apjl, 396, L1
\bibitem[]{} Spergel, D.N. 1994, Warner Prize Lecture, BAAS, 185.7301 
\bibitem[Szapudi \& Szalay 1998]{ss98}  Szapudi, I. \& Szalay, A.S. 1998, 
           \apj, 494, L41 (SS)
\bibitem[]{} Szapudi, I., Prunet, S., Pogosyan, D., Szalay, A.S., \& Bond, J.R.
2001, \apjl, 548, L11 (SPPSB)
\bibitem[]{} Zaldarriaga, M., Spergel, D.N., \& Seljak, U. 1997, \apj, 488, 1 
\bibitem[]{} Wandelt, B. D, 2000, astro-ph/0012416
\bibitem[]{} Wright, E.L., 1996 (astro-ph/9612105)
\bibitem[]{} Wright, E.L., Hinshaw, G. \& Bennett, C.L. 1996, \apj, 458, L53 

\end{thebibliography}
\end{document}